\documentclass[sigconf,natbib]{acmart}
\usepackage{fdiaz}
\usepackage{subfig}
\usepackage{pifont}

\newcommand{\cmark}{\textcolor[rgb]{0.14,0.75,0.00}{\ding{51}}}
\newcommand{\xmark}{\textcolor{gray}{\ding{55}}}

\usepackage{booktabs}

\copyrightyear{2024}
\acmYear{2024}
\setcopyright{rightsretained}
\acmConference[SIGIR-AP '24]{Proceedings of the 2024 Annual International ACM SIGIR Conference on Research and Development in Information Retrieval in the Asia Pacific Region}{December 9--12, 2024}{Tokyo, Japan}
\acmBooktitle{Proceedings of the 2024 Annual International ACM SIGIR Conference on Research and Development in Information Retrieval in the Asia Pacific Region (SIGIR-AP '24), December 9--12, 2024, Tokyo, Japan}\acmDOI{10.1145/3673791.3698428}
\acmISBN{979-8-4007-0724-7/24/12}

\newcommand{\query}[0]{x}
\newcommand{\queries}[0]{\fdset{X}}
\newcommand{\querysamp}[0]{\mathbf{X}}
\newcommand{\querysampsize}[0]{n}
\newcommand{\querydist}[0]{\theta}

\newcommand{\system}[0]{f}
\newcommand{\systemj}[0]{\system'}
\newcommand{\systems}[0]{\fdset{F}}
\newcommand{\systemsamp}[0]{\mathbf{F}}

\newcommand{\metric}[0]{\mu}

\newcommand{\measurements}[0]{\{\metric\}_{\querysamp,\system}}
\newcommand{\measurementsj}[0]{\{\metric\}_{\querysamp,\system'}}
\newcommand{\smeasurements}[0]{\{\metric\}}
\newcommand{\smeasurementsj}[0]{\smeasurements'}
\newcommand{\smetric}[0]{\vec{\mu}}

\newcommand{\aggregate}[0]{\overline{\metric}}
\newcommand{\preference}[0]{\overline{\Delta}}

\newcommand{\risk}[0]{\text{T}}
\newcommand{\srisk}[0]{\overline{\risk}}
\newcommand{\successAtTen}[0]{\text{s@10}}

\newcommand{\gmap}[0]{\text{gavg}}
\newcommand{\auc}[0]{\text{auc}_4}
\newcommand{\gini}[0]{\text{gini}}

\newcommand{\indicator}[0]{\text{I}}

\begin{document}
\title{Pessimistic Evaluation}

\author{Fernando Diaz}
\orcid{0000-0003-2345-1288}
\authornote{work done at Google.}
\affiliation{
\institution{Carnegie Mellon University}
\city{Pittsburgh}
\state{PA}
\country{United States}
}
\email{diazf@acm.org}

\begin{abstract}
Traditional evaluation of information access systems has focused primarily on average utility across a set of information needs (information retrieval) or users (recommender systems).  In this work, we argue that evaluating only with average metric measurements assumes utilitarian values not aligned with traditions of information access based on equal access.  We advocate for pessimistic evaluation of information access systems focusing on worst case utility.  These methods are
\begin{inlinelist}
\item grounded in ethical and pragmatic concepts,
\item theoretically complementary to existing robustness and fairness methods, and
\item empirically validated across a set of retrieval and recommendation tasks.
\end{inlinelist}
These results suggest that pessimistic evaluation should be included in existing experimentation processes to better understand the behavior of systems, especially when concerned with principles of social good.
\end{abstract}



\maketitle
\section{Introduction}
\label{sec:introduction}
Evaluating information access systems that support large populations of users remains a fundamental area of research.  \textit{Individual-level evaluation}, the measurement of the utility that an individual receives from the system in a specific context, has been the focus of work in metric design familiar to the information retrieval community.  \textit{Population-level evaluation} refers to making judgments about a system based on how individual-level utility is distributed across a group or population of users; the most common approach to population-level evaluation is to use the arithmetic mean utility as the measure of system quality.  While individual-level evaluation metrics can often be empirically validated against human preferences, population-level evaluation methods often do not have quantifiable signals that can be used for validation.  Instead, population-level evaluation methods often need to be more rigorously validated for their alignment with normative values about how utility should be distributed across a population.  This is consistent with research in the recommender systems community seeking to incorporate of normative values into system evaluation and design \cite{vrijenhoek:normative-diversity-news,vrijenhoek:radio,normalize-workshop-2023,ferraro:commonality,ferraro:commonality-tors}.  Developing evaluation methods with rigorous conceptual and theoretical foundations is fundamental to translational work between system policy and development.

While the arithmetic mean dominates experimental practice, system design objectives have led to the development of alternative population-level evaluation methods.
Motivated by the need for robust systems, the TREC Robust Track  introduced a number of population-level aggregations capturing how well systems performed on low-performing queries \cite{ROBUST2004}.  As part of the TREC Web Track, the risk-sensitive retrieval task evaluated systems according to how they degraded utility relative to a baseline \cite{kct:trec2013}.  The retrieval efficiency community includes tail efficiency (e.g., 95th percentile query processing time) in evaluations \cite{mackenzie:topk-qps}.
Moreover, recent calls from the recommender system community emphasize evaluation of the distribution of measured metric values over a population \cite{ekstrand:disteval} and from the fairness community to the dis-aggregate metric values across sub-populations \cite{barocas:disaggregation}.

We interrogate average utility and advocate for an alternative population-level evaluation method focused on worst-case analysis, which we refer to as \textit{pessimistic evaluation}.  We ground pessimistic evaluation in existing work in equal information access from the information science community, which is based on a larger body of work in fairness.  This allows pessimistic evaluation to be based on well-justified methods from political theory.  In particular, we introduce the use of lexicographic minimum as a theoretically sound method for pessimistic evaluation.

Our goal is not to demonstrate that pessimistic evaluation dominates existing population-level evaluation based on average system utility.  Rather, we will demonstrate that pessimistic evaluation is
\begin{inlinelist}
\item grounded in ethical and pragmatic concepts,
\item theoretically complementary to existing robustness and fairness methods, and
\item empirically validated across a set of retrieval and recommendation tasks.
\end{inlinelist}
As such, our goal is to demonstrate that pessimistic evaluation complements existing population-level evaluation approaches.
\section{Population-level evaluation}
Population-level evaluation deals with comparing systems given a set of utility  measurements.  In this section, we will introduce the formal problem of population-level evaluation and then discuss how we can compare different population-level evaluation methods.

\subsection{Preliminaries}
For a specific information access task, given an input request $\query$ from the space of all requests $\queries$, a system $
\system$ (from the space of all systems $\systems$) generates an output (e.g., ranking) with a specific utility (e.g., metric value).  An input $\query\in\queries$ can be a text query, as in traditional information retrieval; a user item history, as in recommendation; or a more complex representation such as, for example, a text query combined with user location information and page view history.  For the purpose of our analysis, we only consider measured metric values and are agnostic about whether those utilities are attributed to rankings or strings.  So, although a system generates a  decision (e.g., a ranking, an answer string), we are only interested in the measured utility, computed by an evaluation metric $\metric : \queries\times\systems\rightarrow \Re$  defined as a function over the space of all inputs and systems.  \footnote{We adopt common practice in search evaluation and assume that systems and metrics are deterministic (i.e.,  $\metric(\query,\system)$ will always return the same value).  This means that the measured utility will always be the same for the same request. }

In most situations, we have a distribution $\querydist$ over $\queries$ based on users' engagement with the information access system.  For example, the probability of a certain input may be proportional to the query frequency in a search engine.  In practice, we use a sample of queries $\querysamp\sim\querydist$ to evaluate performance. The shorthand $\measurements = \cup_{\query \in \queries}\metric(\query,\system)$  refers to the set of measured utilities for a system $\system$ over a sample $\querysamp$.  We use $\querysampsize=|\querysamp|$ to refer to the sample size.

\subsection{Problem Definition}
Determining whether a population-level evaluation is appropriate depends on the population-level objectives of the system designer.  In some cases, a quantifiable downstream objective such as revenue can be used to evaluate different population-level evaluation methods.  In many cases, though, population-level objectives reflect less well-defined concepts such as fairness or justice.  We refer to this class of objectives as population-level normative values of a system.

The goal of population-level evaluation is to sort a set of systems $\systemsamp\subseteq\systems$ according to the designers population-level objectives.
Assume that, for each $\system\in\systemsamp$, we have measured utility for $\querysampsize$ inputs.  One can approach this problem by defining an \textit{aggregation function} $\aggregate:\Re^{\querysampsize}\rightarrow\Re$ that reduces a set of measurements into a single scalar value (e.g., an average).  We can then sort $\systemsamp$ according to these aggregate scores.  Alternative, one can define an \textit{order function} $\preference : \Re^{\querysampsize}\times \Re^{\querysampsize}\rightarrow \Re$ that generates a scalar value based on a comparison of two sets of $\querysampsize$ metric measurements.  Note that we can derive an order function from an aggregation function but not the other way around.

\subsection{Desiderata}
Although we can often conduct individual-level metric meta-evaluation by looking at agreement with a ground truth human preference (e.g., `does the metric ordering agree with a user's preference?'), population-level meta-evaluation leans more heavily on  theoretical validation and different methods of empirical validation.  In this study, we consider population-level evaluation desiderata based on criteria from measurement theory \cite{hand:measurement-theory}, previously used in the natural language processing \cite{xiao:nlg-measurement-theory} and retrieval \cite{diaz:lexirecall} communities.  Similar approaches have been used in ranking metric meta-evaluation \cite{valcarce:recsys-ranking-metrics-conf,valcarce:recsys-ranking-metrics-journal}.  In particular, we will validate a population-level evaluation method using the following criteria.

\paragraph{Content validity}  Content validity determines the degree to which the population-level evaluation method is aligned with the theoretical constructs we are interested in measuring.  To assess content validity of a method, we determine whether the method is theoretically consistent with normative values of interest.  For example, if the system designer is interested in high expected utility for users, then computing the average utility over a set of queries would be consistent with this value; averaging the utility for subscribing users only would ignore the expected impact on users with under-performing queries and not be consistent with the system designer's value.

\paragraph{Convergent validity} Convergent validity determines the degree to which a population-level evaluation method is correlated with other methods supporting the same normative values. To assess the convergent validity of a method by, we measure  the empirical correlation between an ordering of $\systemsamp$ by the new approach with orderings of $\systemsamp$ by existing approaches \textit{for  the same normative value}.   For example, if both averaging utility over traffic-weighted queries and averaging utility over unique queries aim to capture the expected impact on users, we can measure the correlation between an ordering of $\systemsamp$ by averaging traffic-weighted queries with a second ordering by averaging unique queries.  Although higher correlation provides evidence that a new measure is consistent with established measures, perfect correlation obviates the need for the new approach.

\paragraph{Discriminant validity} Discriminant validity determines the degree to which a population-level evaluation method is uncorrelated with unrelated methods. To assess the discriminant validity of a method we measure  the empirical correlation between an ordering of $\systemsamp$ by the new approach with orderings of $\systemsamp$ by existing approaches \textit{for  different normative values}.  In this case, we desire \textit{lower} correlation since it provides evidence that a new measure is different from established measures for different constructs.

\paragraph{Sensitivity} Sensitivity refers to how well a population-level evaluation method can distinguish pairs of systems.  A method that is theoretically sound but not sensitive will not be useful in practical settings.  We assess the sensitivity of an approach by how often it is unable to distinguish a pair of systems (i.e., the number of tied systems).
\section{Properties of Population-Level Evaluation}
\label{sec:foundations}
Since information access systems are tools used by people and a metric measurement reflects the utility of a tool to an individual person, each approach for population-level evaluation makes assumptions about the relative importance of some information needs or people compared to others.  As such, we can interpret specific normative values about population-level evaluation as reflecting specific social values.  This is consistent with perspectives  to information access in information science  based on distributive justice \cite{mathiesen:informational-justice} and  echoes recent calls from the recommender system community to reason about the distribution of measured metric values over a population \cite{ekstrand:disteval} and from the fairness community to the dis-aggregate metric values across sub-populations \cite{barocas:disaggregation}.  More generally, the recommender systems community is increasingly exploring the incorporation of normative values into design \cite{vrijenhoek:normative-diversity-news,vrijenhoek:radio,normalize-workshop-2023,ferraro:commonality,ferraro:commonality-tors}.

The foundation of a population-level evaluation method is a normative statement about the population-level objective of the designer.  While most readers will be familiar with using the arithmetic mean as an aggregation function, in this section, we will introduce three normative statements covering several objectives underlying information access system design.  This list is far from exhaustive and the development of appropriate normative values for information access is an active area of research \cite{kruse:normalize-tutorial}.

\subsection{Pareto Property}
The first property we are interested in is the behavior of a method when a single individual's utility improves.
Given two allocations $\measurements$ and $\measurementsj$ with equal utility for $|\querysamp|-1$ individuals, the Pareto property requires that a method prefer the allocation with the higher utility for the remaining individual \cite{hirose:structure-of-aggregation}.  Considering the situation where the utility of a single individual improves ensures that an evaluation respects each person's wellbeing.  That is,  a population-level evaluation that does \textit{not} satisfy this property would sometimes ignore the utility of an individual, even in situations where that of others is not impacted.
From the perspective of information access, the Pareto property means that, the performance of all other queries or users being equal, if a system improves the performance for a single query or user, then it should be preferred.
The Pareto property is formally defined as,
\begin{align}
\forall \query\in\querysamp, \metric(\query,\system) \geq \metric(\query,\system') \wedge \exists \query\in\querysamp, \metric(\query,\system) > \metric(\query,\system') &\rightarrow \system \succ \system'
\end{align}
For example, while the arithmetic mean of $\measurements$ would satisfy the Pareto property,  the median would not, since improving, say, $\min \measurements$ up to the median would not affect the median (or the ordering of systems).  While simple, theoretical consistency with the Pareto principle ensures that, no matter the condition, we respect the strict benefit to individuals.
\subsection{Average Utilitarianism}
While the Pareto property considers a single individual in isolation, we might alternatively consider the expected utility over all individuals.  Average utilitarianism prefers the allocation where the expected utility is higher and is well-aligned with empirical risk minimization, the foundation of many machine learning methods.  In the context of a commercial information access system, when the measurements are correlated with revenue (or inversely correlated with cost), then average utilitarian is often aligned with cumulative revenue.  That said, average utilitarian decision-making focuses on the performance for a random user.  In this sense, as a result, if the population is structured so that some inputs or groups of inputs are over-represented, they can dominate the decision-making.
Average utilitarianism is formally defined as,
\begin{align}
\frac{1}{|\queries|}\sum_{\query\in\queries} \metric(\query,\system) >\frac{1}{|\queries|}\sum_{\query\in\queries} \metric(\query,\system') &\rightarrow \system \succ \system'
\end{align}
Since we compare systems over the same population, we can drop the multiplicative factors and recover the utilitarian condition, defined as the cumulative utility over the population,
\begin{align}
\sum_{\query\in\queries} \metric(\query,\system) >\sum_{\query\in\queries} \metric(\query,\system') &\rightarrow \system \succ \system'
\end{align}
Therefore, we see that average performance has embedded value of utilitarianism \cite{hurka:average-utilitarianism,sidgwick:methodsofethics}.
Average utilitarianism is an implicit value present in almost every information access experiment but is justified from a very specific philosophical tradition.
\subsection{Difference Principle}
\label{sec:foundations:difference}
While the Pareto property focuses on the difference in utility of an individual when the utility of all other individuals is constant, we often care about the fairness of utility distributed across the population.  In particular, \citeauthor{rawls:theory-of-justice} argues that, when an individual does not know which utility in an allocation they will receive, they will rationally decide to prefer the allocation where their worst outcome (i.e., receiving the least utility) is better \cite{rawls:theory-of-justice}.  This is referred to as the difference principle and underlies many approaches to social justice, including many adopted in the machine learning community \cite{liang:fairness-without-demographics,li:worst-case,gupta:rai-worst-case-games,thams:worst-case-robustness-sets}.  Worst case analysis more generally is useful when evaluating the safety of a system and can provide insight when systems are otherwise difficult to distinguish due to ceiling effects \cite{probst:avg-case-worst-case-safety}.

The difference principle is particularly well-suited for information access evaluation.  In the professional librarian community, codes of ethics often include principles of equal access, which are non-utilitarian in nature \cite{hoffmann:rawls-infosci}.  \citet{britz:information-poverty}, based on  Rawlsian theories of social justice,  advocates for measuring the performance of information access based on those least well-served.
In a recommendation context, \citet{singh:saferec} study worst-case evaluation and optimization to design safe reinforcement learning methods.

The difference principle is formally defined using maximin,
\begin{align}
\min_{\query\in\queries}(\metric(\query,\system)) >\min_{\query\in\queries}(\metric(\query,\system')) &\rightarrow \system \succ \system'\label{eq:min}
\end{align}
Unfortunately, because only the minimum value is used, maximin does not satisfy the Pareto principle.  Moreoever, the practical problem with this approach is that most systems will have a minimum value of zero, either because systems all fail in different ways or because of outlier contexts that no system can perform well with.
Indeed,  83\% of runs  submitted to the TREC 2021 Deep Learning Passage Ranking task were tied when evaluated using the worst case NDCG at 10.  In order to address these issues with maximin, \citet{sen:collective-choice-and-social-welfare} introduced the lexicographic minimum or leximin.  Let $\smetric$ be the measurements $\measurements$ sorted in decreasing order; similarly $\smetric'$ for $\measurementsj$.  The leximin preference is defined as,
\begin{align}
\smetric_{i^*} > \smetric'_{i^*} &\rightarrow \system \succ \system'\label{eq:leximin}
\end{align}
where $i^*$ is the maximum index where $\smetric_i \neq \smetric'_i$.  Leximin is frequently adopted in the machine learning literature in lieu of maximin \cite{diana:lexiopt,abernethy:lexiopt}.

Note that, because it focuses on the single worst utility, the difference principle is not statistical by definition.  We contrast this with average utilitarianism, where, because it focuses on the \textit{expected} individual, methods of statistical inference can be leveraged.

\subsection{Compatibility between properties}
\label{sec:foundations:compatibility}
Although these properties are all desirable for different reasons, they are compatible to different degrees.  It is easy to confirm that (average) utilitarianism satisfies the Pareto property; all other utilities being equal, increasing the utility to one individual will increase both the total and average utility.  On the other hand, by focusing only on the worst case, maximin is indifferent between allocations that would be distinguished by the Pareto principle; if $\measurements=[1,0.9,0.1]$ and $\measurementsj=[1,0.8,0.1]$, then the Pareto property would require that $f\succ f'$ but maximin would be indifferent.  Adopting leximin satisfies both the difference principle and the Pareto property.  Finally,   utilitarianism and the difference principle are incompatible; if $\measurements=[1,0.0,0.0]$ and $\measurementsj=[0.3,0.3,0.3]$, then average utilitarianism would observe $f\succ f'$ but satisfying the difference principles would observe $f\prec f'$

The relationship between these properties clarifies what is valued  when we adopt either average utilitarianism or the difference principle.  If we adopt average utilitarianism, we cannot provide guarantees about satisfying Rawlsian fairness; if we adopt the difference principle, we cannot provide guarantees about improving average utility.  This is important to ensure that  evaluation decisions align with normative values and organizational principles.  In cases where the information access provider is maximizing engagement, utilitarianism may be more appropriate than the difference principle.  In cases where the information access provider seeks to provide equal access, the worst case performance may be more important to consider.

\begin{figure*}
\centering
\hfill
\includegraphics[width = 2.25in]{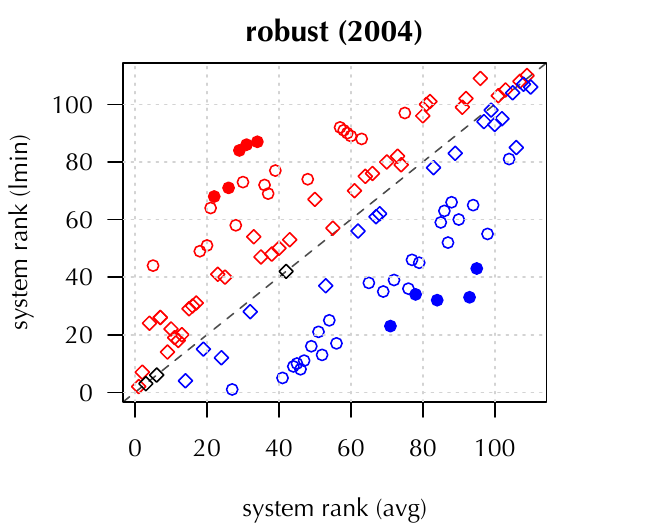}\hfill
\includegraphics[width = 2.25in]{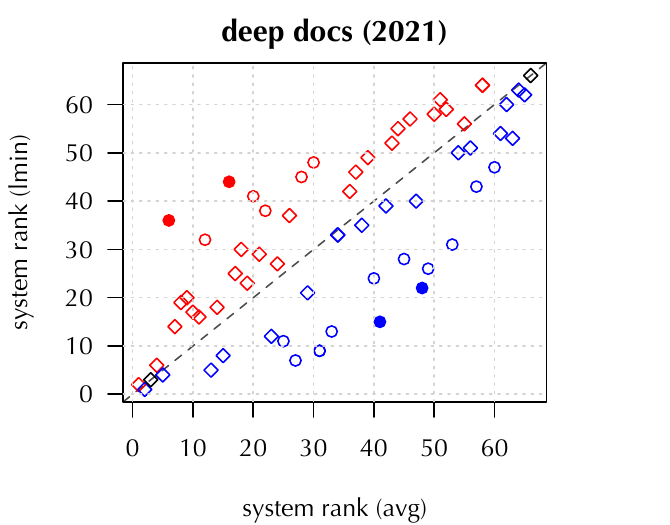}\hfill
\includegraphics[width = 2.25in]{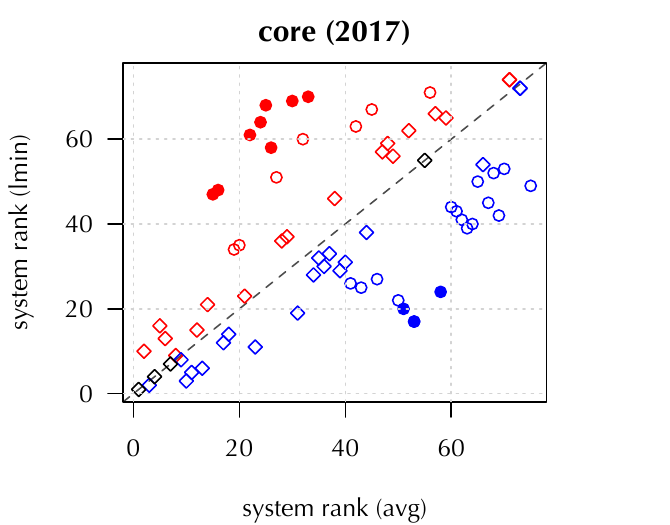}
\hfill
\caption{Ranking of runs in three TREC tracks (see Table \ref{tab:data}) according to associated metrics when ordering systems by average performance (horizontal) and leximin (vertical).  $\diamond$: change in position less than one quintile.  $\circ$: change in position between one and two quintiles.  $\bullet$: change in position greater than two quintiles. Red: degradation in ranking.  Blue: improvement in rank position.  Black: no change in rank position. }\label{fig:vis}
\end{figure*}

In order to compare the empirical ordering by average utilitarianism and by the difference principle, in Figure \ref{fig:vis}, we show the rank position of systems evaluated from the arithmetic mean performance to leximin ordering for several datasets.  Among a number of small adjustments in rank position, several runs degrade from high positions to very low positions (e.g., $6\rightarrow 36$, $12\rightarrow 32$, $16\rightarrow 44$) and from low positions to very high positions (e.g., $25\rightarrow 11$, $31\rightarrow 9$).  In general, these results indicate that leximin indeed empirically captures a different phenomenon than arithmetic mean performance.
\section{Pessimistic Evaluation}
\label{sec:metrics}

We wish to study population-level evaluation methods consistent with the difference principle, maximin and leximin. That said, given the importance of evaluation in information access problems, several methods exist for measuring the worst-performing queries, often in the context of robustness.  In this section, we  review these methods.\footnote{We omit the plethora of fair ranking metrics because they focus on fair exposure of providers rather than utility to users.} We  assume a fixed query sample $\querysamp$ for evaluating systems, consistent with current offline testing practice and, for clarity, write $\smeasurements=\measurements$ and similarly $\smeasurementsj=\measurementsj$.

\paragraph{Geometric mean}\citet{ROBUST2004} uses geometric mean utility to emphasize low measured utility.  This is formally defined as an aggregation,
\begin{align}
\gmap(\smeasurements)&=\left(\prod_{\metric\in\smeasurements}\max\left(\epsilon,\metric\right)\right)^{\frac{1}{\querysampsize}}
\end{align}
where $\epsilon$ is a small value avoids degenerate effects when $\metric=0$.\footnote{\citet{ROBUST2004} sets this value to 0.00001, which we adopt in our experiments.}

\paragraph{Area under the lower  quartile}\citet{ROBUST2004} also uses the area under the curve defined by the lowest quartile of queries.  This is defined as an aggregation,
\begin{align}
\auc(\smeasurements)&=\frac{1}{k}\sum_{j=1}^{k}\frac{1}{j}\sum_{i=1}^{j}\smetric_{\querysampsize-i+1}
\end{align}
where $k=\lfloor\querysampsize/4\rfloor$.

\paragraph{Success at Ten}\citet{ROBUST2004} also uses the fraction of queries that have at least one relevant item in the top ten items and is defined as an aggregation,
\begin{align}
\successAtTen(\smeasurements)&=\frac{1}{\querysampsize}\sum_{\metric\in\smeasurements}\indicator(\metric>0)
\end{align}
where $\metric$ is precision at ten.

\paragraph{Risk-Penalized Gain}
\citet{wang:robust-ranking} uses methods from financial risk modeling to emphasize lower-performing queries.  This method compares a baseline allocation to a treatment allocation, measuring average degradation in  paired difference in performance.  Given two systems $\system$ and  $\system'$, we define the gain of $\system$ over $\systemj$ as an ordering function,
\begin{align}
\risk(\smeasurements,\smeasurementsj,\alpha)&=\frac{1}{\querysampsize}\sum_{\query \in \querysamp}\max\left(0,\metric(\query,\system)-\metric(\query,\systemj)\right)\nonumber\\
&-\frac{1+\alpha}{\querysampsize}\sum_{\query \in \querysamp}\max\left(0,\metric(\query,\systemj)-\metric(\query,\system)\right)
\end{align}
where $\alpha\geq 0$ and we recover ordering by the arithmetic mean when $\alpha=0$.  From a preference perspective, the gain is asymmetric, meaning that $\risk(\smeasurements,\smeasurementsj,\alpha)\neq-\risk(\smeasurementsj,\smeasurements,\alpha)$.  To address this, we combine the two directions as $\srisk(\smeasurements,\smeasurementsj,\alpha)=\risk(\smeasurements,\smeasurementsj,\alpha)-\risk(\smeasurementsj,\smeasurements,\alpha)$.

\paragraph{Gini Coefficient}
The Gini coefficient is a measure of the average difference between all pairs of utilities in an allocation.  Although not used for population-level evaluation in the ranking literature,\footnote{The Gini coefficient \textit{has} been used to measure the inequality of retrieval of items in both search \cite{azzopardi:retrievability} and recommendation \cite{lazovich:disparate-outcomes,do:gini}. } the Gini coefficient is often used in the economics literature to quantify inequity in distribution of utility and is defined as an aggregation,
\begin{align}
\gini(\smeasurements)&=\frac{1}{2\querysampsize^2}\sum_{\metric,\metric' \in \smeasurements}\frac{|\metric-\metric'|}{\overline{\smeasurements}}
\end{align}
where $\overline{\smeasurements}$ is the arithmetic mean of $\smeasurements$.  The Gini coefficient is bounded between 0 and 1, with 0 reflecting maximum equality and 1 reflecting maximum inequality. Note that the Gini coefficient \textit{only} measures inequality and does not capture utility of an allocation.
\section{Theoretical Analysis}
\label{sec:theory}
Understanding whether a particular population-level evaluation method is consistent with a property in Section \ref{sec:foundations} is important for several reasons.  First, because information access systems are increasingly subject to regulation, grounding evaluation in clear conceptual terms provides transparency and theoretical guarantees about the alignment with philosophical and legal principles (e.g., fairness) \cite{dinola:recsys-eu-reg}.  The value of transparency  and these theoretical guarantees extends beyond external regulatory agencies to internal stakeholders, including engineering and policy teams within organization \cite{deshpande:rai-stakeholders}.  Second, theoretically characterization extends axiomatic analysis of individual-level evaluation  to  population-level evaluation \cite{busin:axiometrics,amigo:diversity,parapar:diversity}.

As noted in Section \ref{sec:foundations:compatibility}, although both maximin and leximin satisfy the difference principle by construction, only leximin satisfies the Pareto principle.  In this section, we will examine whether the methods in Section \ref{sec:metrics} are consistent with the properties in Section \ref{sec:foundations}.

\paragraph{Geometric mean}
The geometric mean is equivalent to applying a logarithmic transformation to measurements before computing the arithmetic mean. Because the logarithm is monotonically increasing in utility, if $\smeasurements$ and $\smeasurementsj$ only differ in a single individual who has higher utility in $\smeasurementsj$, then the arithmetic mean of the logarithms will be greater for $\smeasurementsj$.  Next, although equivalent to the arithmetic mean of the logarithms, this does not ensure that the geometric mean is consistent with average utilitarianism.  Consider $\smeasurements=[1, 0.9, 0.1]$ and $\smeasurementsj=[0.5, 0.5, 0.5]$ where the geometric mean prefers $\smeasurementsj$ while the arithmetic mean prefers $\smeasurements$.  Moreover, despite emphasizing lower performance, the geometric mean is not consistent with the difference principle.  If $\smeasurements=[0.25, 0.25, 0.25]$ and $\smeasurementsj=[1, 0.9, 0.1]$, the geometric mean prefers $\smeasurementsj$ while the difference principle would select $\smeasurements$.

\paragraph{Area under the lower  quartile}  Because the area under the lower  quartile only considers a subset of utility values, any change in utility occurs above the lower quartile is ignored.  This means that it does not satisfy the Pareto property.  This also means that it is not consistent with average utilitarianism.  Finally, because the area under the curve accumulates averages within the bottom quartile, situations can arise which are inconsistent with the difference principle.  For example, consider $\smeasurements=[1,0.9, 0.7,0.6, 0.4,0.3, 0.1, 0.05]$ and $\smeasurementsj=[1,0.9, 0.7,0.6, 0.4,0.3, 0.3, 0.0]$, where the area under the lower quartile prefers $\smeasurementsj$ while the difference principle would select $\smeasurements$.

\paragraph{Success at Ten}  While perhaps correlated with many ranking measures, success at ten does not provide any guarantee about utility when considering other measures that inspect beyond the tenth position.  If utility is defined by query-level success at ten, then the population-level success at ten reduces to the arithmetic mean, satisfying the Pareto property and, trivially, average utilitarianism.  Because query-level success at ten is binary, if population-level success at ten detects a difference between two systems, it will be equivalent to the preference detected by leximin, satisfying the difference principle.  That said, for other metrics, population-level success at ten provides no guarantees.

\paragraph{Risk-Penalized Gain}  The symmetric version of gain, $\srisk$, reduces to the difference in arithmetic mean performance with a multiplicative factor of $2+\alpha$, indicating that it satisfies the Pareto property and average utilitarianism but not the difference principle.

\paragraph{Gini Coefficient}
Because the Gini coefficient focuses on measuring equality, it prefers allocations that are more uniform, resulting in inconsistency with the Pareto property.  For example, if $\smeasurements=[0.6,0.5,0.5]$ and $\smeasurementsj=[0.5,0.5,0.5]$, the Gini coefficient would prefer $\smeasurementsj$ while the Pareto property would lead to $\smeasurements$ being preferred.  The same example can be used to demonstrate that the Gini coefficient is not consistent with average utilitarianism.  We can see that Gini coefficient is also inconsistent with the difference principle by considering $\smeasurements=[0.8, 0.6, 0.5, 0.3]$ and $\smeasurementsj=[0.5, 0.3, 0.3, 0.2]$, where the Gini coefficient would prefer $\smeasurementsj$ while the difference principle would prefer $\smeasurements$.

\paragraph{Summary} We summarize results in Table \ref{tab:eval-properties}.  These analyses all compromise the content validity of these population-level evaluation methods from the perspective of the properties discussed in Section \ref{sec:foundations}.  This does not suggest that the methods are not useful in evaluating systems, but that they are inconsistent with various properties important to designing information access systems.   In  Section \ref{sec:experiments}, we will move from theoretical analyses to study the empirical behavior of these methods.
\begin{table}
\centering
\caption{Summary of population-level evaluation properties: Pareto property (PP), average utilitarianism (AU), difference principle (DP) for minimum (min), leximin  (lmin), arithmetic mean (avg), geometric mean  (gavg), success at ten (s@10), area under the lower quartile ($\auc$), risk-penalized gain (gain).   }\label{tab:eval-properties}
\begin{tabular}{lccc}
\hline
&PP & AU & DP \\
\hline
min  &\xmark &\xmark& \cmark  \\
lmin  & \cmark & \xmark&\cmark  \\
avg & \cmark & \cmark &\xmark\\
gavg & \cmark & \xmark&\xmark\\
s@10 & \xmark & \xmark &\xmark\\
$\auc$& \xmark & \xmark&\xmark\\
gain & \cmark & \cmark&\xmark\\
gini& \xmark & \xmark&\xmark\\
\hline
\end{tabular}
\end{table}
\section{Empirical Analysis}
\label{sec:experiments}
Having examined the properties of several population-level evaluation methods, we now turn to studying the empirical relationship between these methods and leximin.  Our objective in this section is to measure the convergent validity, discriminant validity, and sensitivity of leximin.  In order to measure convergent validity, we compute Kendall's $\tau$ between the ranking of systems ordered by leximin and the ranking of systems ordered by other pessimistic evaluation methods (i.e., those in Section \ref{sec:metrics}).  We adopt $\tau_b$ to handle ties \cite{kendall:tau-b}.  In order to measure discriminant validity, we compute Kendall's $\tau$ between  the ranking of systems ordered by leximin and the ranking of systems ordered by other methods \textit{not} focused on pessimistic evaluation.  In this case, we consider the arithmetic mean and leximax, which is analogous to leximin but starts at the best-case utility.  In order to measure sensitivity, we count the number of tied systems under each population-level evaluation method.

\subsection{Data}
\label{sec:experiments:data}
We analyzed population-level evaluation across a wide range of information access tasks where using the difference principle is well-motivated (Section \ref{sec:foundations}).  For  information retrieval contexts, we used official runs submitted to 24 different TREC tracks.  For recommendation contexts, we used publicly available runs for three recommendation tasks: movielens, beerAdvocate, and libraryThing \cite{valcarce:recsys-ranking-metrics-journal}.  Dataset details are available in Appendix \ref{app:data}.

We focused analysis on comparing the following population-level methods: minimum, leximin, arithmetic mean, geometric mean, success at ten, and lowest quartile area under the curve.  We omit symmetric risk-penalized gain because it is equivalent to the arithmetic mean and the Gini coefficient because it does not capture total utility.

\subsection{Results}
\label{sec:experiments:results}
Results for corpora using average precision as the utility measure are presented in Table \ref{tab:bigtable}. Results for other metrics on the robust (2004) corpus are presented in Table \ref{tab:bigtable-by-metric}.

When measuring convergent validity, we are interested in higher correlation with population-level methods intended to capture the same higher level concept.  In Table \ref{tab:bigtable}, when comparing leximin to minimum, geometric mean, success at ten, and lowest quartile area under the curve, we observe $\tau$ values in general ranging from 0.50 to 0.90, suggesting high empirical agreement.  Cases where correlations are weaker occur when the population-level evaluation method incurs substantial ties (e.g., web (2009), deep-docs (2020), legal (2007)).  Although leximin and minimum both capture the difference principle, we observe a correlation lower than 1 because leximin computes preferences even when we observe a tie in minimum utilities.  The next highest correlation after minimum is with the geometric mean, which is the only other pessimistic evaluation method that satisfies the Pareto property.  This means that, unlike success at ten and lowest quartile area under the curve, like leximin, it considers the full set of queries.

When measuring discriminant validity, we are interested in lower correlation with population-level methods intended to capture the different higher level concepts, in our case average utilitarianism and best-case utility.  When comparing leximin to arithmetic mean, we observe $\tau$ values in general in the range from 0.20 to 0.60, suggesting low to moderate correlation.  Comparing the $\tau$ across methods for a single condition (i.e., row), we see that the correlation with the arithmetic mean is lower than the correlations with pessimistic methods.  Moreover, the correlation is consistently weakest with leximax, which captures the best-case performance and should be low.

Table \ref{tab:bigtable-by-metric} demonstrates that our observations about convergent and discriminant validity are consistent across other metrics.  We noticed that, especially for metrics with rank cutoffs (e.g., ndcg@10, p@10), the success at ten method correlated much higher with leximin.

Finally, when inspecting the number of ties for each method, leximin always is tied for the lowest value and is comparable to other methods with high sensitivity.  This is especially salient  when comparing leximin with minimum, the only other method consistent with the difference principle, which consistently demonstrates tied performance. As such, if we are interested in satisfying the difference principle and conducting experiments, leximin should be preferred to minimum.

\begin{table*}
\caption{Kendall's $\tau$ for various datasets using average precision as the utility metric.  Comparison between system ranking based on leximin  (lmin) and minimum score  (min), geometric mean  (gavg), success at ten (s@10), area under the lower quartile ($\auc$), arithmetic mean (avg), and leximax (lmax).  Number of ties in parentheses.  Dashes reflect situations where all systems are tied according to the method.  Bold: highest $\tau$ in convergent validity analysis.  Italics: lowest correlation in discriminant validity analysis. }\label{tab:bigtable}
{\small
\begin{tabular}{lc|c@{\hskip 0.25em}c|c@{\hskip 0.25em}c@{\hskip 2em}c@{\hskip 0.25em}c@{\hskip 2em}c@{\hskip 0.25em}c@{\hskip 2em}c@{\hskip 0.25em}c|c@{\hskip 0.25em}c@{\hskip 2em}c@{\hskip 0.25em}c}
&&\multicolumn{10}{c|}{pessimistic evaluation}&\multicolumn{4}{c}{} \\
& nruns & \multicolumn{2}{c|}{lmin} & \multicolumn{2}{c@{\hskip 2em}}{min} & \multicolumn{2}{c@{\hskip 2em}}{gavg} & \multicolumn{2}{c@{\hskip 2em}}{s@10} & \multicolumn{2}{c@{\hskip 2em}|}{$\auc$} & \multicolumn{2}{c@{\hskip 2em}}{avg} & \multicolumn{2}{c}{lmax} \\
\hline
robust (2004) & 110 & 1.000 & (2) & \textbf{0.882} & (54) & 0.665 & (2) & 0.566 & (95) & 0.682 & (2) & 0.474 & (2) & \textit{0.193} & (2) \\
&&&&&&&&&&&&&&\\
core (2017) & 75 & 1.000 & (6) & \textbf{0.998} & (16) & 0.557 & (6) & 0.481 & (73) & 0.592 & (6) & 0.433 & (6) & \textit{0.331} & (6) \\
core (2018) & 72 & 1.000 & (2) & \textbf{0.987} & (16) & 0.629 & (2) & 0.603 & (67) & 0.664 & (9) & 0.512 & (2) & \textit{0.450} & (2) \\
&&&&&&&&&&&&&&\\
web (2009) & 48 & 1.000 & (0) & 0.287 & (46) & 0.761 & (0) & 0.666 & (32) & \textbf{0.785} & (5) & 0.475 & (0) & \textit{0.043} & (0) \\
web (2010) & 32 & 1.000 & (2) & \textbf{0.970} & (12) & 0.556 & (2) & 0.523 & (28) & 0.661 & (2) & 0.244 & (2) & \textit{0.071} & (2) \\
web (2011) & 61 & 1.000 & (8) & \textbf{0.863} & (39) & 0.733 & (8) & \textit{0.372} & (57) & 0.671 & (8) & 0.640 & (8) & 0.444 & (8) \\
web (2012) & 48 & 1.000 & (4) & \textbf{0.957} & (22) & 0.586 & (4) & 0.321 & (42) & 0.648 & (4) & 0.401 & (4) & \textit{0.302} & (4) \\
web (2013) & 61 & 1.000 & (8) & \textbf{0.974} & (30) & 0.564 & (8) & 0.506 & (57) & 0.708 & (8) & 0.370 & (8) & \textit{0.151} & (8) \\
web (2014) & 30 & 1.000 & (4) & \textbf{0.992} & (10) & 0.473 & (4) & 0.508 & (26) & 0.621 & (4) & 0.386 & (4) & \textit{0.095} & (4) \\
&&&&&&&&&&&&&&\\
deep-docs (2019) & 38 & 1.000 & (0) & \textbf{0.967} & (10) & 0.366 & (0) & 0.527 & (36) & 0.616 & (0) & 0.260 & (0) & \textit{-0.084} & (0) \\
deep-docs (2020) & 64 & 1.000 & (0) & 0.539 & (54) & \textbf{0.619} & (0) & 0.545 & (62) & 0.563 & (0) & 0.408 & (0) & \textit{0.277} & (0) \\
deep-docs (2021) & 66 & 1.000 & (6) & \textbf{0.975} & (21) & 0.520 & (6) & 0.641 & (65) & 0.604 & (6) & 0.210 & (6) & \textit{-0.041} & (6) \\
deep-docs (2022) & 42 & 1.000 & (0) & \textbf{0.883} & (20) & 0.847 & (0) & 0.697 & (37) & 0.844 & (0) & 0.784 & (0) & \textit{0.526} & (0) \\
deep-docs (2023) & 5 & 1.000 & (0) & \textbf{0.949} & (2) & 0.600 & (0) & 0.527 & (2) & 0.600 & (0) & 0.600 & (0) & \textit{0.400} & (0) \\
&&&&&&&&&&&&&&\\
deep-pass (2019) & 37 & 1.000 & (0) & 0.549 & (31) & \textbf{0.628} & (0) & 0.563 & (35) & 0.532 & (0) & 0.580 & (0) & \textit{0.517} & (0) \\
deep-pass (2020) & 59 & 1.000 & (0) & \textbf{0.808} & (35) & 0.615 & (0) & 0.523 & (49) & 0.617 & (3) & 0.501 & (0) & \textit{0.416} & (0) \\
deep-pass (2021) & 63 & 1.000 & (0) & 0.589 & (51) & \textbf{0.704} & (0) & 0.519 & (57) & 0.642 & (2) & 0.520 & (0) & \textit{0.051} & (0) \\
deep-pass (2022) & 100 & 1.000 & (0) & \textbf{0.793} & (66) & 0.731 & (0) & 0.632 & (92) & 0.722 & (0) & 0.649 & (0) & \textit{0.457} & (0) \\
deep-pass (2023) & 35 & 1.000 & (0) & \textbf{0.824} & (22) & 0.771 & (0) & 0.755 & (27) & 0.768 & (0) & 0.681 & (0) & \textit{0.382} & (0) \\
&&&&&&&&&&&&&&\\
legal (2006) & 34 & 1.000 & (0) & - & (34) & \textbf{0.722} & (0) & 0.433 & (25) & 0.712 & (6) & 0.490 & (0) & \textit{0.005} & (0) \\
legal (2007) & 68 & 1.000 & (0) & \textbf{0.943} & (25) & 0.514 & (0) & \textit{0.323} & (60) & 0.433 & (3) & 0.430 & (0) & 0.347 & (0) \\
&&&&&&&&&&&&&&\\
podcasts (2020) & 14 & 1.000 & (0) & 0.711 & (10) & \textbf{0.912} & (0) & 0.769 & (7) & 0.739 & (4) & 0.868 & (0) & \textit{0.560} & (0) \\
podcasts (2021) & 27 & 1.000 & (11) & 0.649 & (21) & 0.725 & (11) & \textbf{0.770} & (21) & 0.704 & (12) & 0.642 & (11) & \textit{0.487} & (11) \\
&&&&&&&&&&&&&&\\
tot (2023) & 30 & 1.000 & (2) & - & (30) & \textbf{0.714} & (2) & 0.414 & (8) & 0.503 & (26) & 0.341 & (2) & \textit{0.304} & (2) \\
&&&&&&&&&&&&&&\\
movielens & 21 & 1.000 & (0) & - & (21) & 0.752 & (0) & 0.790 & (0) & \textbf{0.802} & (4) & 0.676 & (0) & \textit{0.600} & (0) \\
beerAdvocate & 21 & 1.000 & (0) & - & (21) & \textbf{0.857} & (0) & 0.781 & (0) & - & (21) & 0.771 & (0) & \textit{0.705} & (0) \\
libraryThing & 21 & 1.000 & (0) & - & (21) & \textbf{0.952} & (0) & 0.933 & (0) & 0.900 & (8) & 0.914 & (0) & \textit{0.838} & (0)
\end{tabular}
}
\end{table*}

\begin{table*}
\caption{Kendall's $\tau$ for robust (2004) using R-Precision (rp), average precision (ap), normalized discounted cumulative gain (ndcg), precision (p), and reciprocal rank (rr).  Formatting identical to Table \ref{tab:bigtable}. }\label{tab:bigtable-by-metric}
{\small
\begin{tabular}{l|c@{\hskip 0.25em}c|c@{\hskip 0.25em}c@{\hskip 2em}c@{\hskip 0.25em}c@{\hskip 2em}c@{\hskip 0.25em}c@{\hskip 2em}c@{\hskip 0.25em}c|c@{\hskip 0.25em}c@{\hskip 2em}c@{\hskip 0.25em}c}
&\multicolumn{10}{c|}{pessimistic evaluation}&\multicolumn{4}{c}{} \\
& \multicolumn{2}{c|}{lmin} & \multicolumn{2}{c@{\hskip 2em}}{min} & \multicolumn{2}{c@{\hskip 2em}}{gavg} & \multicolumn{2}{c@{\hskip 2em}}{s@10} & \multicolumn{2}{c@{\hskip 2em}|}{$\auc$} & \multicolumn{2}{c@{\hskip 2em}}{avg} & \multicolumn{2}{c}{lmax} \\
\hline
rp & 1.000 & (2) & - & (110) & 0.840 & (2) & 0.778 & (95) & \textbf{0.848} & (4) & 0.516 & (2) & \textit{0.192} & (2) \\
ap & 1.000 & (2) & \textbf{0.882} & (54) & 0.665 & (2) & 0.566 & (95) & 0.682 & (2) & 0.474 & (2) & \textit{0.193} & (2) \\
ndcg & 1.000 & (2) & \textbf{0.882} & (58) & 0.687 & (2) & 0.526 & (95) & 0.723 & (2) & 0.524 & (2) & \textit{0.178} & (2) \\
&&&&&&&&&&&&&\\
ndcg@100 & 1.000 & (2) & 0.352 & (103) & \textbf{0.802} & (2) & 0.688 & (95) & 0.787 & (2) & 0.494 & (2) & \textit{0.233} & (2) \\
p@100 & 1.000 & (2) & 0.347 & (110) & 0.788 & (2) & 0.688 & (95) & \textbf{0.853} & (4) & 0.471 & (2) & \textit{0.005} & (2) \\
&&&&&&&&&&&&&\\
ndcg@10 & 1.000 & (2) & - & (110) & 0.890 & (2) & \textbf{0.987} & (95) & 0.908 & (4) & 0.530 & (2) & \textit{0.141} & (2) \\
p@10 & 1.000 & (2) & - & (110) & 0.909 & (2) & \textbf{0.987} & (95) & 0.962 & (41) & 0.526 & (2) & \textit{0.174} & (2) \\
&&&&&&&&&&&&&\\
rr & 1.000 & (2) & \textbf{0.882} & (64) & 0.703 & (2) & 0.589 & (95) & 0.660 & (2) & 0.527 & (2) & \textit{0.472} & (2)
\end{tabular}
}
\end{table*}

\section{Discussion}
\label{sec:discussion}
We were motivated to rigorously define and understand pessimistic evaluation from basic concepts grounded in normative principles of fairness and information access.  Our theoretical results demonstrate that leximin is the only population-level evaluation in our suite that satisfies the fundamental Pareto property and difference principle.  Our empirical results demonstrate the convergent and discriminant validity of leximin while also having high sensitivity.

We based our adoption of the difference principle on work from information science advocating for Rawlsian fairness in information access \cite{britz:information-poverty,hoffmann:rawls-infosci} and separate work in the machine learning community \cite{liang:fairness-without-demographics,li:worst-case,gupta:rai-worst-case-games,thams:worst-case-robustness-sets}.  As such, providing formal guarantees of population-level methods aligning with it is important, especially in responsible artificial intelligence contexts where stakeholders from multiple disciplines need to align \cite{deshpande:rai-stakeholders}.

Although we have grounded pessimistic evaluation in arguments from information science, alternative positions may demand alternative population-level evaluation methods.  Just as average utilitarianism can be justified for commercial organizations, other properties can be justified by other contexts.  For example, in some commercial  situations, there is a privileged group of subscribed customers that deserves substantially more attention during population-level evaluation to ensure retention \cite{wu:ltv-tutorial}.  Or, alternative notions of justice may require entirely new population-level fairness evaluation formalisms \cite{green:escaping-impossibility-of-fairness}.

As discussed in Section \ref{sec:foundations:difference}, leximin is not statistical by definition since it emphasizes the absolute worst case instead of the expected case.    This does not imply that there is not uncertainty in estimating worst-case performance differences since it can  arise from query sampling, randomness in system decisions, in addition to other sources.

\begin{figure}
\centering
\includegraphics[width = 2.25in]{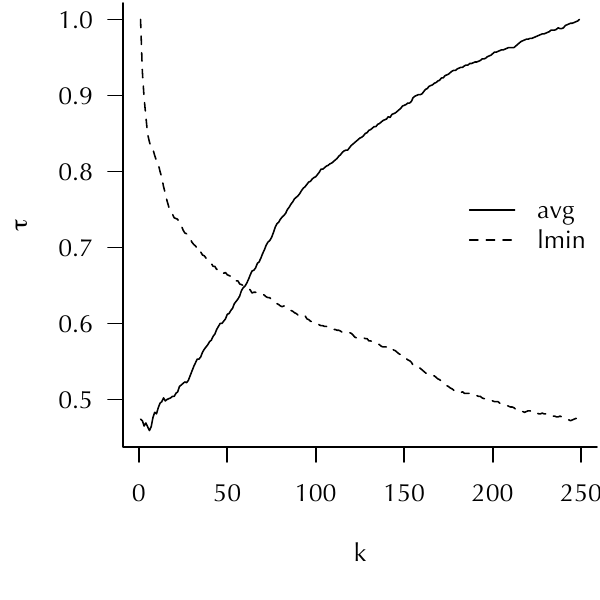}
\caption{Kendall's $\tau$ between  system orderings by smoothed leximin and arithmetic mean (solid) and leximin (dashed) using average precision on Robust 2004 (see Section \ref{sec:experiments:data} for details). }\label{fig:smoothing}
\end{figure}

One way to introduce uncertainty is to define a relaxed version of leximin.  For example, we can use a moving average to smooth measurements.  For a lag of $k$, we compare $\sum_{j=0}^{k-1}\smetric_{i+j}$ and $\sum_{j=0}^{k-1}\smetric'_{i+j}$ instead of $\smetric_{i}$ and $\smetric'_{i}$.  This means $i$ iterates from $n-(k-1)$ to 1, recovering the arithmetic mean when $k=n$.  In Figure \ref{fig:smoothing}, we demonstrate how adjusting $k$ generates rankings of systems smoothly transitioning between the difference principle (low values of $k$) and average utilitarianism (high values of $k$).  This behavior is similar to lower quartile area under the curve except that, like the leximin, it backs off to higher quantiles in the presence of a tie.
Smoothed leximin satisfies the Pareto property because, if $\smeasurements$ and $\smeasurementsj$ only differ in a single individual who has higher utility in $\smeasurementsj$, then each $\sum_{j=0}^{k-1}\smetric_{i+j}\leq\sum_{j=0}^{k-1}\smetric'_{i+j}$ and at least one inequality is strict.
Moreoever, it is easy to see that, unless $k=\querysampsize$, smoothed leximin evaluation can be inconsistent with the average utilitarian decision.  As an example, consider $\smeasurements=[1,0.0,0.0]$ and $\smeasurementsj=[0.2,0.2,0.2]$ where, unless $k=\querysampsize$, smoothed leximin prefers $\smeasurementsj$.  Similarly, unless $k=1$, smoothed leximin evaluation can be inconsistent with the difference principle.  As an example, consider $\smeasurements=[0.1,0.1,0.1]$ and $\smeasurementsj=[0.25,0.25,0.0]$ where, unless $k=1$, smoothed leximin prefers $\smeasurementsj$. We believe smoothed leximin provides one way to begin to explore statistical methods for pessimistic evaluation.

\begin{figure}
\centering
\subfloat[]{\includegraphics[width = 1.6in]{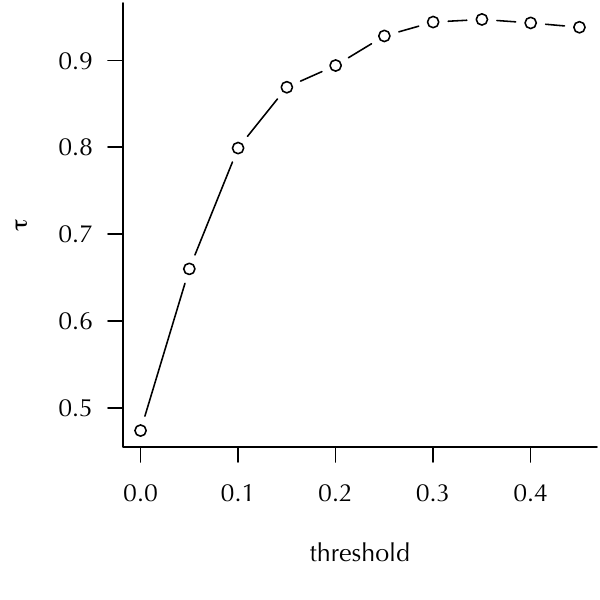}}\hfill
\subfloat[]{\includegraphics[width = 1.6in]{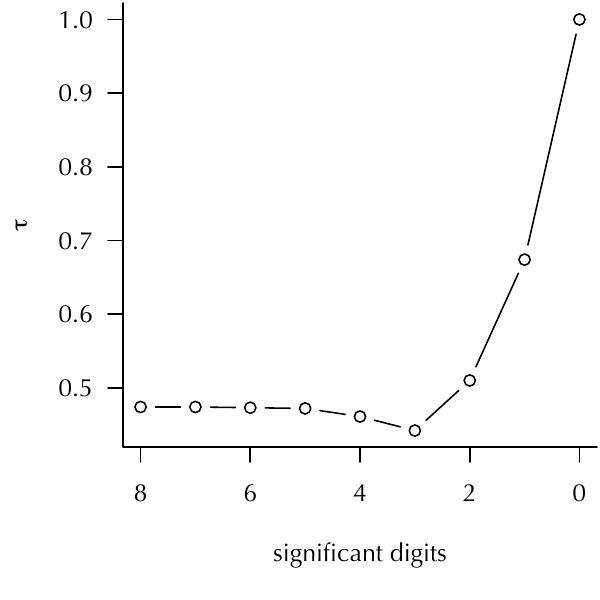}}
\caption{Discretization of average precision values for Robust 2004. (a) Metric values below the threshold set to zero.  (b) Metric values quantized to a specific number of significant digits.   }\label{fig:discretization}
\end{figure}

While leximin provides a lens into worst-case performance, there are situations when aggregation-based methods are more appropriate.  For example,  we may be interested in a single score assigned to each system for a downstream process or decision.  Aggregation-based methods may also obviate the need for leximin in some metric conditions.  For example, leximin becomes increasingly similar to the arithmetic mean as the number of discrete utility values decreases.  When there are two values---as with success at ten---leximin and the arithmetic mean are equivalent.  To see how discretization of utility values affects the relationship between leximin and the arithmetic mean, we conducted the following experiment.  Under the first discretization method, we progressively set all utility valued below a threshold to 0.  Under the second discretization method, we removed significant digits from the utility value.   Figure \ref{fig:discretization} shows the Kendall's $\tau$ between the leximin ordering and the arithmetic mean ordering as we discretized average precision values. Under both discretization methods, we observe a gradual convergence with the arithmetic mean.  This suggests that, even if we are interested in the difference principle, there are situations where the arithmetic mean is equivalent.

At a higher level, we hope that a rigorous conceptual and theoretical grounding of population-level evaluation provides a template for others interested in developing new methods.  In particular, because population-level evaluation often explicitly or implicitly captures social values, understanding those foundations when analyzing existing or designing new methods is important for policy, legal, and ethical integrity.
\section{Conclusion}
\label{sec:conclusion}
While the information retrieval community has a rich history of  individual-level evaluation research focused around metric design, substantially less work exists studying population-level evaluation, even though it is fundamental to all retrieval experiments.  As an example of the importance of analysis of population-level evaluation, we introduced pessimistic evaluation through leximin, a method firmly grounded in robust philosophical and moral traditions.  We contrast the guarantees provided by this theoretical foundation with related methods from robustness measures in information retrieval.  We further demonstrated content, convergent, and discriminant validity of leximin as  well as a competitive sensitivity.  We advocate information retrieval experimenters---especially those in organizers with values aligned with the difference principle---to complement existing population-level methods with leximin.
\appendix

\section{Data}
\label{app:data}
All TREC runs and relevance judgments were downloaded from NIST.\footnote{\href{https://trec.nist.gov/results.html}{https://trec.nist.gov/results.html}}  Recommendation runs and judgments were downloaded from a public repository.\footnote{\href{https://github.com/dvalcarce/evalMetrics}{https://github.com/dvalcarce/evalMetrics}}  Datasets are detailed in Table \ref{tab:data}.  Metrics were computed using the official trec\_eval package.\footnote{\href{https://github.com/usnistgov/trec_eval}{https://github.com/usnistgov/trec\_eval}}
\begin{table}[t]
\caption{Datasets used in empirical analysis.  Runs submitted to the associated TREC track or recommendation task.  }\label{tab:data}
{\small
\begin{tabular}{lcccc}
\hline
&   requests    &   runs    &   rel/request &   docs/request\\
\hline
robust (2004) &       249 &  110 &    69.93  &    913.82 \\
\\
core (2017)     &       50  &  75   &   180.04 &        8853.11 \\
core (2018)     &       50  &  72   &   78.96  &        7102.61 \\
\\
web (2009)      &       50  &  48 & 129.98  &   925.31 \\
web (2010)      &       48  &  32 & 187.63  &   7013.21 \\
web (2011)      &       50  &  61 & 167.56  &   8325.07 \\
web (2012)      &       50  &  48 & 187.36  &   6719.53 \\
web (2013)      &       50  &  61 & 182.42  &   7174.38 \\
web (2014)      &       50  &  30 & 212.58  &   6313.98 \\
\\
deep-docs (2019)        &       43  &  38 & 153.42 & 623.77 \\
deep-docs (2020)        &       45  &  64 & 39.27  & 99.55 \\
deep-docs (2021)        &       57  &  66 & 189.63  & 98.83 \\
deep-docs (2022)        &       76  &  42 & 1245.62  &  100\\
deep-docs (2023)        &       82  & 5  &  75.10 &  100\\
\\
deep-pass (2019)        &       43  &  37 & 95.40  & 892.51 \\
deep-pass (2020)        &       54  &  59 & 66.78  & 978.01 \\
deep-pass (2021)        &       53  &  63 & 191.96  & 99.95 \\
deep-pass (2022)        &       76  &  100 & 1315.22  &  100\\
deep-pass (2023)        &       82  &  35 & 103.18  & 100 \\
\\
legal (2006)     &       39  &  34   &   110.85 &       4835.07\\
legal (2007)     &       43  &  68   &   101.02  &     22240.30\\
\\
podcasts (2020)      &       48  &  14 & 43.67  &   963.40 \\
podcasts (2021)      &       50  &  27 & 30.80  &   781.15 \\
\\
tot (2023)      &       150  &  31 & 1  &   1000 \\
\\
movielens   &       6005    &  21 & 18.87  & 100.00\\
libraryThing &   7227    &   21  &   13.15   &  100.00\\
beerAdvocate &   17564    &   21  &   13.66   & 99.39\\
\hline
\end{tabular}
}
\end{table}
\bibliographystyle{ACM-Reference-Format}
\balance
\bibliography{XX-references.bib}
\end{document}